\documentclass[9pt, twocolumn, twoside]{osajnl}
\journal{ao}
\setboolean{shortarticle}{false}

\title{Spectroscopic study of a diffusion-bonded sapphire cell for hot metal vapors}

\author[1]{Naota Sekiguchi}
\author[1]{Takumi Sato}
\author[2]{Kiyoshi Ishikawa}
\author[1,*]{Atsushi Hatakeyama}

\affil[1]{Department of Applied Physics, Tokyo University of Agriculture and Technology, Koganei, Tokyo 184-8588, Japan}
\affil[2]{Graduate School of Material Science, University of Hyogo, Kamigori,  Hyogo 678-1297, Japan}
\affil[*]{Corresponding author: hatakeya@cc.tuat.ac.jp}

\dates{Compiled \today}
\ociscodes{(020.0020) Atomic and molecular physics; (020.3690) Line shapes and shifts; (020.2070) Effects of collisions; (120.0120) Instrumentation, measurement, and metrology; (300.6210) Spectroscopy, atomic}
\doi{}

\begin{abstract}
Characteristics of a diffusion-bonded sapphire cell for optical experiments with hot metal vapors were investigated. The sapphire cell consisted of sapphire-crystal plates and a borosilicate-glass tube, which were bonded to each other by diffusion bonding without any binders or glues. The glass tube was attached to a vacuum manifold using the standard method applied in glass processing, filled with a small amount of Rb metal by chasing with a torch, and then sealed. The cell was baked at high temperatures and optical experiments were then performed using rubidium atoms at room temperature. The sapphire cell was found to be vacuum tight, at least up to 350$^{\circ}$C, and the sapphire walls remained clear over all temperatures. From the optical experiments, the generation of a background gas was indicated after baking at 200$^{\circ}$C. The background gas pressure was low enough to avoid pressure broadening of absorption lines but high enough to cause velocity-changing collisions of Rb atoms. The generated gas pressure decreased at higher temperatures, probably due to chemical reactions.
\end{abstract}

\setboolean{displaycopyright}{true}

\begin{document}

\maketitle

\section{Introduction}

Alkali-metal-vapor cells made of silica-based glasses are used in atomic physics in a range of experiments such as hyperfine atomic clocks \cite{Camparo2007}, optical magnetometers \cite{BudkerRomalis2007}, atomic gyroscopes \cite{Kornack2005PRL}, and noble-gas hyperpolarization \cite{Walker97}. Some experiments with alkali-vapor cells require high temperatures to increase the number density of alkali atoms; however, there is an upper temperature limit of approximately 200$^{\circ}$C when using glass cells in optical experiments, since the glasses become mechanically weak and discolored due to absorption of chemically-reactive alkali metals at high temperatures. Although aluminosilicate glass can withstand alkali metal attack better than borosilicate and silica glasses, aggressive alkalis can discolor aluminosilicate to an opaque brown after a few weeks of use at $\sim$350$^{\circ}$C \cite{Babcock2005}.

Aluminum oxide materials (Al$_{2}$O$_{3}$) such as crystalline sapphire and polycrystalline alumina (PCA) are alkali-resistant even at higher temperatures (for example, up to $\sim$1500$^{\circ}$C for sodium \cite{Schlejen1987} 
and $\sim$1000$^{\circ}$C for cesium \cite{Sarkisyan2005RSI}) unless they are exposed to hot lithium vapor \cite{Brog1967, Slabinski1971, Ishikawa2016}. Since sapphire crystal is transparent across a wide range of wavelengths, from the visible to the infrared spectrum, sapphire crystals have been used as optical windows in a number of alkali-vapor cell experiments conducted at high temperatures \cite{Arthur1980, Schlejen1987, Gallagher1995, Laliotis2008, Cundiff2008, Keaveney2012PRL, Shmavonyan2015, Pichler2016}. Sapphire-windowed cells have also been used in atom-surface interaction experiments \cite{Failache1999PRL, Laliotis2008, Keaveney2012PRL, Shmavonyan2015} and to store hyperpolarized noble gases and liquids \cite{PinesBudker2004PRL, Masuda2005}. The sapphire cell reported in \cite{Masuda2005} has the unique feature that the sapphire windows were diffusion-bonded to a sapphire tube without any binders or glues: the sapphire-sapphire bonding reduces stress-induced birefringence of the sapphire windows caused by differences in expansion curves \cite{Sarkisyan2005RSI}; the absence of glues may result in low contamination and low background-gas pressure. However, the characteristics of diffusion-bonded sapphire cells containing alkali atoms at high temperatures have not been investigated.

In this work, the characteristics of a diffusion-bonded sapphire cell for optical experiments at high temperatures were investigated. Sapphire single-crystal plates were bonded with each other by diffusion bonding, and a borosilicate-glass tube was attached so that metal samples could be vacuum sealed by flaming the glass tube, using a similar method to that applied for all-glass cells. The sapphire cell containing rubidium (Rb) atoms was repeatedly characterized using optical measurements based on the absorption lines and velocity-changing collisions (VCCs) of the Rb atoms at room temperature, after exposure to high temperatures. It was confirmed that the sapphire cell was vacuum tight and transparent, at least up to 350$^{\circ}$C. Production of a background gas was indicated at 200$^{\circ}$C by modifications in the Rb absorption lines and an increase in VCC rate. The background gas production decreased at 250$^{\circ}$C, indicating that the gas was fixed on the walls by some chemical reaction.

\section{Sapphire cell}

\begin{figure}[htbp]
\centering
\includegraphics[clip, width=70mm]{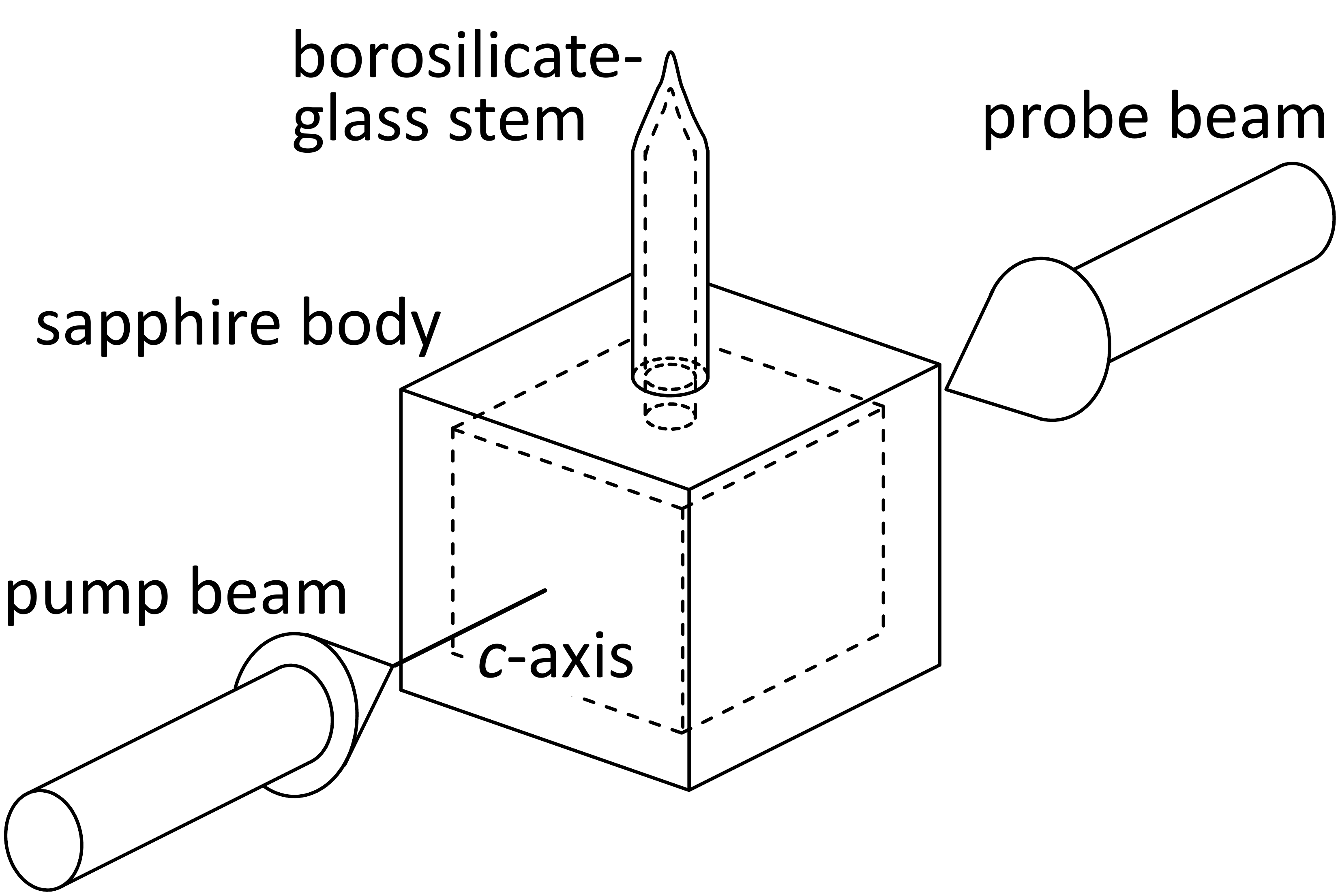}
\caption{The sapphire cell (inner size, $14 \times 14 \times 14$~mm$^3$) and optical configuration. The pump and probe beams propagate collinearly along the $c$-axis of the sapphire crystal.
}
\label{sapphire_cell}
\end{figure}

\begin{figure}[htbp]
\centering
\includegraphics[clip, width=35mm, bb=0 0 72 105]{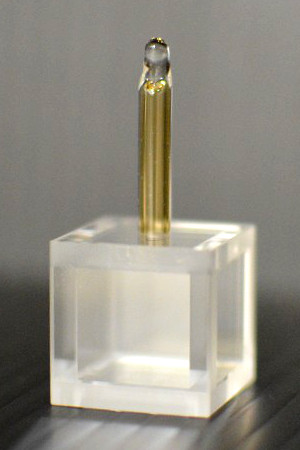}
\caption{Photograph of the cell after the heating process at 400$^{\circ}$C. The sapphire body is clear, although the glass tube becomes discolored.  
}
\label{photo}
\end{figure}

The alkali-vapor cell had a cubic shape made of sapphire plates and a borosilicate-glass tube (Japan Cell Co., Ltd.). These elements were bonded with each other by diffusion bonding without binders or glues, as shown in Fig.~\ref{sapphire_cell}. The optic axes of the bonded crystal walls were matched with each other, similar to a hollowed out bulk single crystal. The contact area of the glass tube (outer diameter $\phi$4.0~mm, inner diameter $\phi$2.4~mm) with the crystal was small but had adequate mechanical strength. 
The diffusion bonding process, conducted at approximately 1400$^{\circ}$C also gave rise to annealing, that is, the reconstruction of the polished surface led to a decrease in atom adsorption at the step edges \cite{Sarkisyan2005RSI}. The glass tube used for the stem (thermal expansion coefficient: $6.4 \times 10^{-6}${$^{\circ}$C$^{-1}$}) belonged to a high-expansion group of the borosilicate glasses \cite{VacuumDevices} and was designed to approximate the thermal expansion of crystalline sapphire across a wide temperature range. As described below, in a heat-proofing test with Rb metal, the glass tube was brown colored but the contact remained vacuum tight up to 350$^{\circ}$C.

The above-described manufacturing processes were performed in air until the sapphire cell was attached to a vacuum manifold. The procedure for vacuum sealing was similar to that used for all glass alkali-vapor cells. The sapphire cell was evacuated at a baking temperature of 400$^{\circ}$C for a whole day. Rb metal at natural isotope ratio was transferred from a metal reservoir to the cell using a gas torch at room temperature, and the sapphire cell was sealed by flaming the glass stem at a background pressure lower than $3 \times 10^{-5}$~Pa. Most of the metal was placed in the glass stem and a thin film spot was visible on the sapphire wall facing the glass stem. The finished cell was kept at room temperature for six months, and the background gas pressure was then confirmed to be less than the detection limit by the optical measurements described below. For the heat-proofing test, the cell was placed at a certain aging temperature $T_{\rm A}$ for six hours and then returned to room temperature for the optical measurements. After the first set of aging and measurement processes, subsequent sets was performed at higher temperatures, using aging temperatures from 50$^{\circ}$C to 400$^{\circ}$C in 50$^{\circ}$C increments. Since the whole cell was uniformly heated, the glass stem gradually took on a brown coloration as the test proceeded, but the sapphire windows remained clear for the whole life of the cell (see Fig.~\ref{photo}). The Rb metal in the stem gradually disappeared as the test proceeded. After the last aging period at 400$^{\circ}$C, no atomic spectral lines were detected, probably because the metal was all absorbed into the glass at this temperature.  

\section{Optical experiments}

During optical experiments the sapphire cell was placed in a permalloy magnetic shield at room temperature and immersed in a longitudinal magnetic field of $40$~$\mu$T. As shown in Fig.~\ref{sapphire_cell}, pump and probe lights derived from a laser diode were collinearly introduced to the sapphire cell along the $c$-axis of the sapphire crystals. The optical frequency of the laser diode was tuned to the $^{85}$Rb $D_{2}$ lines of the ground-state hyperfine level $|F=3\rangle$. The beam diameters at $1/e^{2}$ intensity of the pump and probe lights were 0.54~mm and 6.4~mm, respectively. The powers and polarizations of the laser beams were changed depending on the type of optical measurement.

The first optical measurement was saturation spectroscopy. The laser frequency was scanned over the $D_{2}$ lines of $^{85}$Rb, ${|F=3\rangle \rightarrow |F' \rangle}$. A pump light of 10~$\mu$W and a probe light of 5~$\mu$W were linearly polarized perpendicularly to each other. The intense pump light excited Rb atoms from the ground level $|F=3\rangle$ to the $|F'\rangle$ levels, leading to a reduction in the absorption of the probe light. The pump light amplitude was modulated at $\nu_{\rm m} = 100$~kHz in this measurement, and the absorption reduction was measured by lock-in detection. In addition, the pump light created a hyperfine polarization and a population imbalance between the hyperfine ground levels, through the deexcitation $|F' = 2, 3\rangle \rightarrow |F=2\rangle$. The hyperfine polarization also reduced the absorption of the probe light, but the lock-in detection had poor sensitivity to the hyperfine polarization, because its relaxation time was longer than the modulation frequency $\nu_{\rm m}$. However, the probe light detected the hyperfine polarization as a modification to the saturation absorption spectrum at high VCC rates \cite{Aminoff1982JP,Bhamre2013PRA}. Modifications such as line broadenings and an elevation in the tails of absorption lines, known as a pedestal, will be expected for VCC rates comparable to $2\pi \nu_\mathrm{m} \sim 6\times 10^{5}~\mathrm{s^{-1}}$.

The second optical measurement was measurement of the VCC rate of Rb atoms with the background gas \cite{Sekiguchi2016}. The laser frequency was tuned to the transition ${|F=3\rangle \rightarrow |F'=4 \rangle}$. The pump and the probe powers were 500~$\mu$W and 10~$\mu$W, respectively, and the pump light was picked up with an acousto-optic modulator as a circularly polarized pulse with a duration of 1~$\mu$s. Atoms moving perpendicularly to the optical axis were selectively polarized by the pump light from the Maxwell-Boltzmann distribution, due to the Doppler effect. The circularly-polarized probe light passed through the spin-polarized atoms and was absorbed. The absorption of the probe light depended on the direction of circular polarization. We measured the velocity-selective spin-polarization by taking the difference in absorption of each light polarization $\sigma_{\pm}$; at low VCC rates, the absorption difference decayed, mainly due to atom transit across the probe beam, and the decay was accelerated when the movement direction was changed by VCCs with the background gas. We were able to detect the VCC rate on the order of $10^{4}$~s$^{-1}$ under these experimental conditions. A detailed discussion of VCC rate measurement has been provided in Ref.~\cite{Sekiguchi2016}. Spin relaxations due to collisions with background gas were considered to be negligible in this work.

\begin{figure}[htbp]
\centering
\includegraphics[clip, width=75mm]{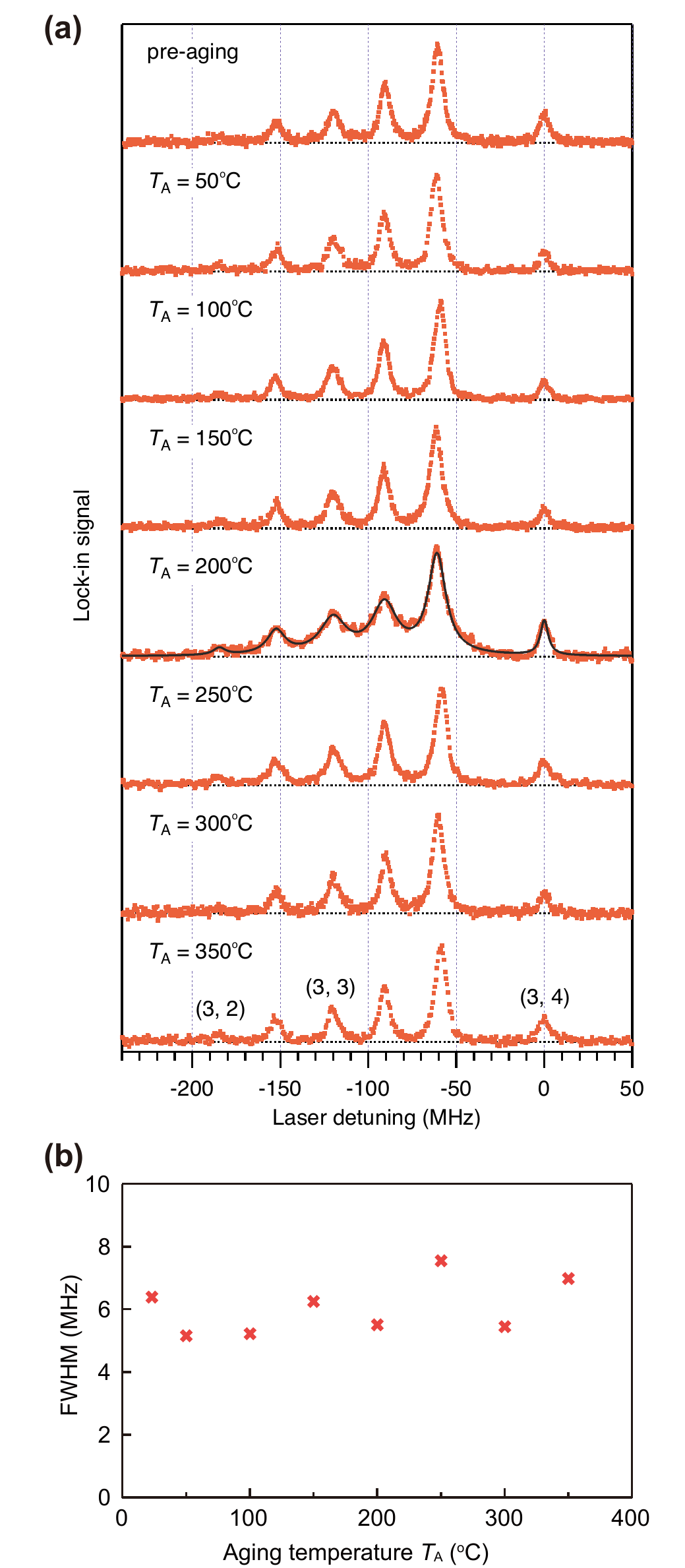}
\caption{(a) Normalized saturation absorption spectra measured at room temperature before aging (pre-aging) and after each aging process at temperature $T_{\rm A}$ as a function of laser detuning from the $|F=3 \rangle \rightarrow |F'=4 \rangle$ transition. The peaks corresponding to the transition between hyperfine levels $|F=3 \rangle \rightarrow |F' \rangle$ are indicated as (3, $F'$) and the crossover peaks are observed at the midpoints of the transition frequencies. The solid curve shows an example of fitting by Lorentz functions for $T_{\rm A} = 200^{\circ}$C. (b) Full-width at half-maximum (FWHM) of the (3,4) peak. The width before aging is presented at 23$^{\circ}$C.  
}
\label{saturation}
\end{figure}

\begin{figure}[htbp]
\includegraphics[clip, width=85mm]{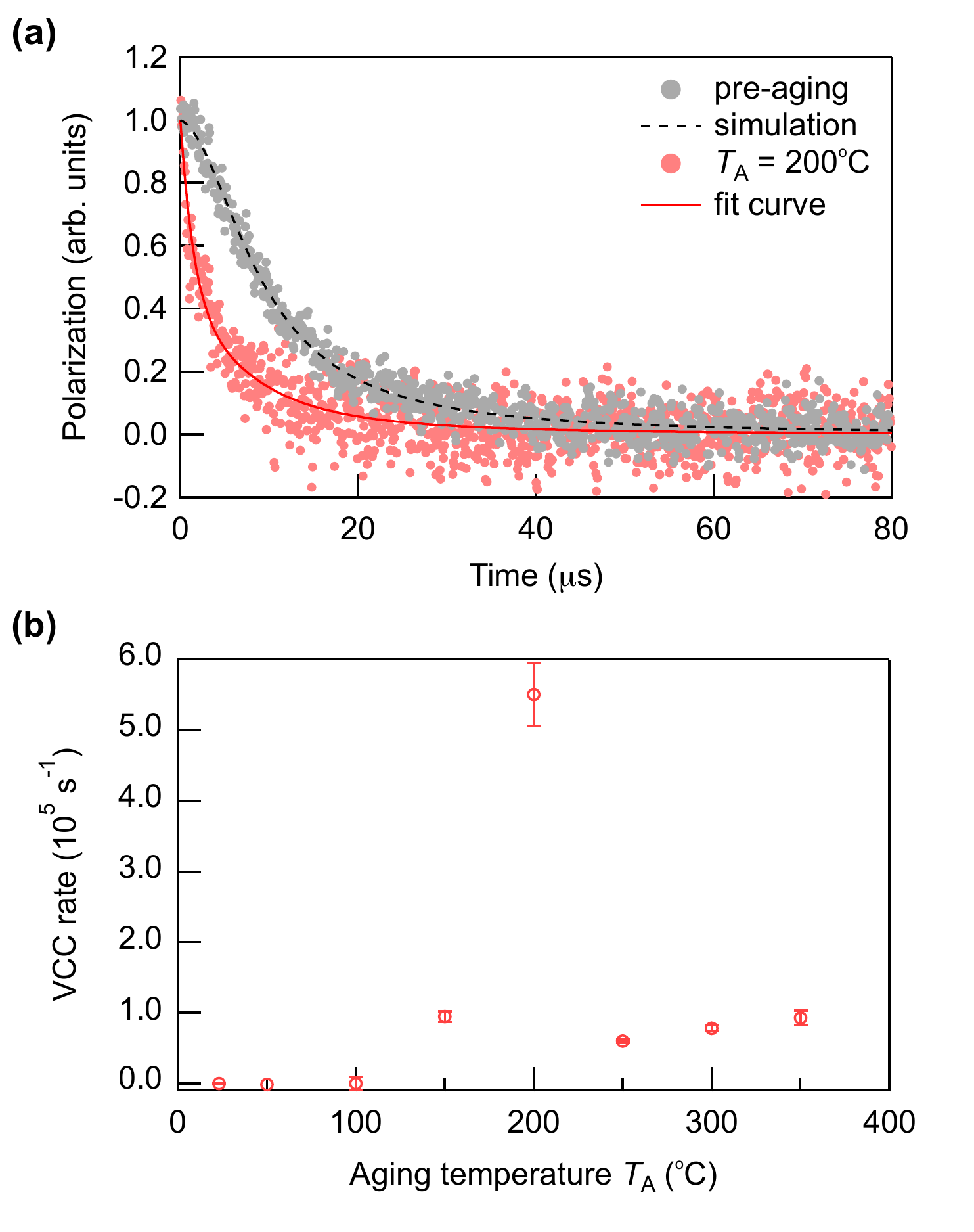}
\caption{(a) The transient signal in the VCC rate measurement at room temperature before aging (pre-aging) and after aging at $T_{\rm A}=200^{\circ}$C. The broken curve shows the simulated time behavior assuming no VCCs. The fit curve using the VCC rate as a fitting parameter is shown by the solid curve. The signals were normalized to the time 0~s. (b) VCC rate after each aging process. 
}
\label{VCC}
\end{figure}

\section{Aging characteristics at high temperatures}
First, for the heat-proofing test, optical measurements were conducted prior to heating to characterize the initial state of the cell. The saturation absorption spectrum for the pre-aging cell is shown in Fig.~\ref{saturation}a. The linewidth of the peak corresponding to the $(F, F')=(3,4)$ transition was measured to be 6.4~MHz and mainly of the natural width, ${\Delta \nu_{\mathrm{n}} = 6.1~\mathrm{MHz}}$. The time evolution of the signal in the VCC rate measurement for the pre-aging cell is shown in Fig.~\ref{VCC}a. The measured signal agrees well with the simulation with no VCCs; therefore, the concentration of the background gas in the initial cell was below the detection limit.

By performing the heat-proofing test, the saturation spectra for the aged sapphire cell were modified as shown in Fig.~\ref{saturation}a. We fitted a Lorentzian function to the (3, 4) peak and derived the linewidth of the peak. The linewidths of the (3, 4) peaks were almost the same over the test as shown in Fig.~\ref{saturation}b, although the other peaks were modified. It is likely that the modifications were caused by redistributions of the hyperfine polarization in velocity space due to VCCs with the background gas. The hyperfine polarization is produced through the (3, 2) and (3, 3) transitions but not through the (3, 4) transition; therefore, the modifications were not pronounced at the (3, 4) peak. The modifications reached a maximum at an aging temperature of 200$^{\circ}$C and decreased at the next aging temperature.

The variations in VCC rate were also observed by the VCC rate measurement as shown in Fig.~\ref{VCC}. The red points in Fig.~\ref{VCC}a show the time transient signal after aging at 200$^{\circ}$C. The decay of the signal was clearly faster than that for the pre-aging signal. To find the VCC rate, the signals were fitted by a theoretical model with a strong-collision assumption \cite{Sekiguchi2016}, where the atomic velocity is thermalized by single collisions with the background gas. 
More accurately, collision kernels should be used for intermediate collisions \cite{Marsland2012PRA,McGuyer2012PRL,Happer09Book,Keilson1952QAM}. 
Figure~\ref{VCC}b shows the VCC rates over the heat-proofing test. As with the saturation absorption spectra, the VCC rates increased by up to $5.5 \times10^{5}~\mathrm{s^{-1}}$, 
which roughly corresponds to a pressure on the order of 1~Pa \cite{Sekiguchi2016}, 
after aging at 200$^{\circ}$C aging and decreased at higher aging temperatures. The background gas, therefore, was not caused by a permanent leak. It is likely that the temporarily generated gas was fixed on the walls by chemical reactions at the higher temperature.

The optical experiments showed that background-gas pressure was increased by aging at $T_{\rm A} \lesssim 200^{\circ}$C. The gas can be produced by chemical reaction of Rb metal on the glass and sapphire surfaces. 
To remove hydrophobic groups, the sapphire plates were cleaned with alcohol, alkaline solution, and deionized water, and the glass tube with alcohol, alkaline and acid solutions, and deionized water. 
Adsorbed molecules such as H$_2$O, CO$_2$, N$_2$, and noble gases should be desorbed by diffusion bonding at 1400$^{\circ}$C in air and by baking at 400$^{\circ}$C in vacuum. Nonetheless, hydrophilic groups may be present, for example, {$-$OH} on aluminum atoms and {$-$H} on oxygen atoms of the sapphire crystal. Plausible molecules produced by chemical reaction were H$_2$, O$_2$, and H$_2$O. Because the gas pressure was sufficiently high to change the velocity distribution of the polarized atoms, chemicals that are highly reactive with alkali metals, O$_2$ and H$_2$O, can be eliminated from the list of possibilities. Hydrogen molecules react with Rb metal to form the hydride RbH at high temperature \cite{Ishikawa2007PRL}, and this is consistent with the result that the gas pressure decreased after aging at $T_{\rm A} \gtrsim 200^{\circ}$C. Based on the discussion above, it is likely that the temporarily generated gas was molecular hydrogen.

\section{Summary}
In this work, we confirmed the capability of a diffusion-bonded sapphire cell for optical experiments at high temperatures. The sapphire cell consisted of sapphire plates and a borosilicate-glass tube, which were diffusion bonded with each other without any binders or glues. The glass tube enabled sealing of metal samples using the standard method applied in glass processing. Optical measurements for Rb atoms in the sapphire cell were conducted at room temperature after heating processes at high temperatures. It was confirmed that the diffusion bonding of the sapphire plates and the borosilicate-glass tube was vacuum tight over a long period of time and that the sapphire walls remained clear while the glass tube took on a brown coloration over the heating processes. A background gas was produced by heating with the Rb metal at a temperature of $\lesssim 200~^{\circ}$C, and the gas pressure decreased in the sealed cell at higher temperatures. A plausible gas was hydrogen. No optical signals were detected after aging at 400$^{\circ}$C, probably because the metal was absorbed into the glass material. Since some change in the thermal expansion coefficient of the glass is expected after absorption of alkali metals, a heat-proofing test of the diffusion bonding is required with a large amount of metal at high temperatures. It is possible to prevent metal transfer from the sapphire body to the glass tube, in the opposite direction to that observed in glass cells \cite{Karaulanov2009}, with a sapphire ball plugged in the through-hole. Accordingly, by sealing a sufficient quantity of alkali metal in the sapphire body, a dense atomic vapor will be optically measured under a homogeneous temperature with clear windows. This work demonstrates that sapphire cells could also be applied to optical study of high temperature chemical reactions. Once the glass tube has absorbed all the metal, it can be manufacturer-replaced for reuse of the expensive sapphire body.

This work was supported in part by JSPS KAKENHI Grant Numbers JP25610115 and JP16H04030.  
N.S. is Research Fellow of Japan Society for the Promotion of Science.

\end{document}